\documentclass[12pt]{iopart}
\usepackage{amssymb}
\begin{document}
\renewcommand{\baselinestretch}{1.125}
\parskip 3pt plus.2pt minus.1pt
\def\e{{\rm e}}
\def\Rop{{\mathbb R}}
\def\metric#1{g^{\raise1pt\hbox{$\scriptstyle\rm#1$}}_{\mu\nu}}
\def\goesas{\mathop{\sim}\limits}
\def\w#1{\,\hbox{#1}}\def\etal{{\em et al.}}
\def\kms{\w{km}\;\w{sec}^{-1}}\def\kmsMpc{\kms\w{Mpc}^{-1}}
\def\hm{\,h^{-1}\hbox{Mpc}}
\def\beq{\begin{equation}} \def\eeq{\end{equation}}
\def\PRL#1{Phys.\ Rev.\ Lett.\ {\bf#1}} \def\PR#1#2{Phys.\ Rev.\ #1 {\bf#2}}
\def\ApJ#1{Astrophys.\ J.\ {\bf#1}} \def\PL#1#2{Phys.\ Lett.\ #1 {\bf#2}}
\def\MNRAS#1{Mon.\ Not.\ R.\ Astr.\ Soc.\ {\bf#1}}
\def\CQG#1{Class.\ Quantum Grav.\ {\bf#1}}
\def\GRG#1{Gen.\ Relativ.\ Grav.\ {\bf#1}}
\def\LCDM{$\Lambda$CDM}
\def\FE{field equations}
\title{What is General Relativity?}
\author{Alan A.~Coley\;\dag\;
and David L. Wiltshire\;\P\;}
\address{\dag\; Department of Mathematics and Statistics, Dalhousie University,
B3H 3J5 Halifax, Canada}
\address{\P\; Department of Physics and Astronomy, University of Canterbury,
Private Bag 4800, Christchurch 8140, New Zealand}
\eads{\mailto{aac@mathstat.dal.ca},
\mailto{david.wiltshire@canterbury.ac.nz}}

\begin{abstract}
General relativity is a set of physical and geometric principles, which
lead to a set of (Einstein) field equations that determine the gravitational
field, and to the geodesic equations that describe light propagation and the
motion of particles on the background. But open questions remain,
including: What is the scale on which matter and geometry are
dynamically coupled in the Einstein equations?
Are the field
equations valid on small and large scales? What is the largest
scale on which matter can be coarse grained while following a geodesic
of a solution to Einstein's equations? We address these questions.
If the field equations are causal evolution equations, whose average on
cosmological scales is not an exact solution of the Einstein equations,
then some simplifying physical principle is required to explain the statistical
homogeneity of the late epoch Universe. Such a principle may have its
origin in the dynamical coupling between matter
and geometry at the quantum level in the early Universe.
This possibility is hinted at by diverse approaches to quantum gravity
which find a dynamical reduction to two effective dimensions at high
energies on one hand, and by cosmological observations which are beginning
to strongly restrict the class of viable inflationary phenomenologies on
the other. We suggest that the foundational principles of general relativity
will play a central role in reformulating the theory of spacetime structure to
meet the challenges of cosmology in the 21st century.
\end{abstract}
\bigskip
\leftline{Invited Comment for {\em Physica Scripta} focus issue on
{\em 21st Century Frontiers}}\bigskip \leftline{Physica Scripta {\bf92} (2017)
053001}
\maketitle

\newpage
\section{Introduction}

It is now one hundred years since Einstein formulated his General Theory of
Relativity (or General Relativity or GR for short) in 1915 \cite{GR}. This past
year there have been many articles written to celebrate this unique achievement.
This paper is intended to briefly look at what theory Einstein proposed, what
the theory has actually come to mean now to theoretical physicists,
and what principles of GR have been abandoned in most attempts to
tackle the unsolved challenges of cosmology and quantum gravity.

Public lectures on cosmology often motivate their descriptions of
current theories as being built on the wonderful achievements of GR of the
last century. However, such comments are misleading when they
describe models that dispense with some of the most important innovations of GR,
such as the dynamical coupling of matter and geometry.
The choices we still typically make can be traced back to
the precedents set by Einstein, who first applied GR to the Universe as a whole
\cite{esu} within less than
two years of formulating the theory. Back in 1917, at
a time before the expansion of the Universe was known and when the very
existence of separate galaxies was still a matter of debate, Einstein wrestled
with many of the open foundational questions of his 1915 theory \cite{esu}.
The choices he made in dealing with those open questions were informed by the
observational knowledge and philosophical preconceptions of his day.

On the centenary of relativistic cosmology, it is important to
ask to what extent current approaches to the theoretical frontiers of quantum gravity and cosmology
still accurately reflect the essence of Einstein's 1915 theory.
In light of the enormous developments of both quantum theory and
observations of the Universe over the past 100 years, could we approach
the open challenges in fresh ways while remaining true to the
foundational principles of GR?

In particular, GR revolutionized our picture of spacetime, changing it from
a fixed stage on which dynamics is played out to a relational
structure in which the dynamics of geometry and matter are inextricably
linked. As Wheeler famously stated: {\em``Matter tells
space how to curve, and space tells matter how to move.''} Yet standard
cosmology as currently practised could be better
summarized as: {\em``Friedmann tells space how to curve, and Newton tells
matter how to move.''} The standard
$\Lambda$ Cold Dark Matter (\LCDM) cosmology assumes an average
Friedmann-Lema\^{\i}tre-Robertson-Walker (FLRW)
evolution in which space expands to maintain constant average spatial
curvature, while the growth of structure in the regime when
perturbations become non-linear is treated by Newtonian $N$--body
numerical simulations\footnote{Full GR computational cosmology is
an immense technical challenge that has hitherto been simply too daunting.
Nonetheless, in the past year the first investigations have begun
\cite{cc}, opening what we believe will become a major frontier for
GR for decades to come.}.

Although the \LCDM\ cosmology
is extremely successful (up to various possible anomalies and tensions
\cite{lcdm}), it requires sources of energy density---dark energy and
non-baryonic CDM---that dominate the present epoch Universe, while
never having been directly detected. Rather than abandoning the 100--year
old innovations of GR, should we be thinking harder about its central
principles in order to tackle these cosmic mysteries?

In many ways GR is not fully utilized in cosmology and modern physical theories
that grapple with questions of the unification of gravity and gauge
interactions, the very early Universe, dark matter and dark energy.
It is perhaps true that certain approaches are more in the spirit of GR:
e.g., loop quantum gravity (LQG) \cite{LQG} and backreaction in cosmology \cite{BR}.
However, it
is possible that when facing fundamental puzzles at the interface of quantum
gravity and cosmology in the very early Universe, the
foundational principles of GR may
resurface in any future redevelopments that are ultimately successful.

In this article we suggest that GR as Einstein originally envisaged is
not a finished theory. As we point out in the final section, the latest
cosmological observations and many independent lines of investigation in
quantum gravity point to a need to rethink our ideas about the nature and
origin of spacetime structure in ways which demand as much conceptual
creativity and rigorous mathematical innovation
as Einstein himself first applied 100 years ago.

\section{What is General Relativity?}

General relativity is based on a set of physical and geometrical principles:

\begin{itemize}
\item 1. Spacetime structure.

\item 2. Equivalence principle.

\item 3. Local causality.

\item 4. Preferred local coordinate frames.
\end{itemize}
A very concise introduction to these topics for non-experts is given in the Appendix.\footnote{For a somewhat longer non-technical introduction to GR
see, e.g., ref.\
\cite{DebonoSmoot}.}

In item 1 the structure includes the fact that the spacetime is a
4-dimensional differentiable manifold of Lorentzian signature
and that the field equations are generally covariant (tensorial)
The Strong Equivalence Principle leads to the postulate that spacetime is
endowed with a metric
and a pseudo-Riemannian structure. Other physical principles include notions of local energy--momentum conservation and
an appropriate Newtonian limit.
These together lead to:\medskip

\noindent
{\bf{I. The Einstein field equations.}}\medskip

Three aspects of this within GR are:

\begin{itemize}

\item Ia. The \FE\ are hyperbolic partial differential equations, and
do not restrict the topology of spacetime a priori.

\item Ib. The geometry and matter are dynamical.

\item Ic. Energy is not conserved in general. Instead the energy--momentum
tensor is covariantly conserved, dynamically mixing the energy of
matter and geometry.

\end{itemize}

In item 3 local causality includes the fact that GR subsumes special relativity
(SR), in which local freely falling reference
frames play an important role, and the fact that gravitational waves travel at the speed of light.
It is then assumed that freely falling test particles follow timelike geodesics and light follows null geodesics
of the spacetime geometry. This then leads to:\medskip

\noindent
{\bf{II. Particles and photons follow geodesics.}}\medskip

One very important aspect is: {\em{what are the scales on which GR is valid?}}.
Clearly,
GR and the field equations are certainly not valid on small scales when quantum effects
must be taken into consideration. To obtain the classical theory of gravity, in principle one must then
average the theory of Quantum Gravity (QG), and either obtain GR or an alternative.
This leads to generalized classical \FE\ and necessarily a violation of
Ia (loop quantum gravity) or a violation of Ib (string theory).

Perhaps of more interest here is whether GR is valid on large (cosmological)
scales. In a similar way, we must average GR at the classical level
to obtain a coarse grained theory on large (cosmological) scales.
Averaging
or coarse graining may well lead to a generalization of I (modified \FE) but also perhaps to II also.
Most generalizations of GR are simple amendments of the
\FE, but within the same mathematical framework and with similar physical principles.
But this is not always the case with
averaging; although we do expect corrected/modified \FE, in general
the averaged geometry is not necessarily Riemannian (or even metric)
and we must also average geodesics
(which affect the interpretation of motions),
which may lead to changes in the fundamental postulates of the theory.

\section{Small scales: Quantum theory}\label{qm}

Unlike in GR, time and space are
not on an equal footing in quantum mechanics (time is treated classically
whereas space is associated with a quantum description). Many modern theories
of quantum gravity (QG) do not respect Einstein's revolutionary way of
interpreting gravitational physics geometrically \cite{QGr1,QGr2}.
Most approaches to QG are based on a covariant Lorentzian action, which
entails an integral over the manifold plus an integral over the boundary.
This is a global object and is only well defined when the
topology is fixed.
Unlike most field theories, in GR
\cite{QGr1}
the field variable is the spacetime metric, $g_{\mu\nu}$, defined on a
four-dimensional (4D) manifold, $\cal M$. In this case, the natural volume element in the integrals
itself depends on the field variables $g_{\mu\nu}$,
and hence its variation must be
taken into account when calculating functional derivatives.
In the canonical approach a family of spacelike surfaces is introduced and used to
construct a Hamiltonian.

To solve for the correct local evolution (the \FE\
or equations of motion) from the action, we need to know the topology,
appropriate boundary conditions and
(in an open manifold) the conditions at infinity \cite{QGr1}. Therefore,
there needs to exist a preferred (global) timelike vector, and hence
a global topology $\Rop \times {\cal M}^3$, for it to make sense.
Regarding the boundary conditions,
the action certainly
makes more sense in a closed universe (i.e., a compact $S^3$).
The surface integral is more complicated in open universes, in which boundary terms enter in a more fundamental
way (and it is not known in general what these terms should be).
Therefore, there are problems with boundary conditions at infinity
for an open manifold: we need to know where infinity is (definition), and conditions at infinity (which might be
timelike or null).
In an open or closed universe we need to add surface terms on a case-by-case basis
(e.g. different for each type of spacetime).

Therefore, in order to do canonical quantization, especially in the Hamiltonian formulation of QG,
we need to know the topology,
appropriate boundary conditions and
(in an open manifold) the conditions at infinity \cite{AAC}.
In the {\em{canonical}} approach, the decomposition into three spatial dimensions
and one time dimension seems to be contrary to the whole spirit of GR
\cite{QGr1}. This is a valid concern, but successful background independent
approaches to quantum GR \cite{QGr3}
(such as LQG and causal dynamical triangulations) accept this price.

In GR, there is no background geometry. The space-time metric itself is the
fundamental dynamical variable. In the {\em{covariant}} approach the emphasis
is on field-theoretic techniques, and does not necessarily involve a $1+3$
decomposition of space-time but it is background dependent.
In {\em{covariant}} approaches to QG,
such as string theory \cite{GQ}, the spacetime metric
is split into a kinematical background and dynamical fluctuations.
The first
step is to split the space-time metric
$g_{\mu\nu}$ in two parts, $g_{\mu\nu}= \eta_{\mu\nu} +
h_{\mu\nu}$, where $\eta_{\mu\nu}$ is the background
metric, often chosen to be flat, and it is only $h_{\mu\nu}$
that is quantized. That is, it is
assumed that the underlying spacetime can be taken to be a
continuum, endowed with a smooth background geometry, and the
quantum gravitational field can be treated as any other quantum
field evolving on this fixed background.

String theory is not restricted to perturbing about flat spacetime, however.
The key advances
from the 1990s onward have involved non-perturbative results
which extend the scope of string theory to include additional extended
objects such as $D$-branes, dualities
between different higher-dimensional
spacetime vacua in particular limits, and the anti-de Sitter
space/conformal field theory (adS/CFT) correspondence. This led to
a change from treating the spacetime metric as coupling constants
in the quantum field theory of the 2-dimensional string world sheet to its
reconstruction from a holographic, dual theory \cite{QGr2}. In either
viewpoint, the spacetime metric has a geometric interpretation which
is far more complicated than in classical GR.

In many theories of fundamental physics there are
geometric classical corrections to GR.
A Lorentzian spacetime with global topology  $\Rop \times {\cal M}^3$ is
completely classified
by its set of scalar polynomial curvature invariants \cite{invariants}.
Thus, in the canonical approach to QG all gravitational degrees of freedom are
curvature invariants.
However, a Lorentzian degenerate Kundt spacetime\footnote{A Lorentzian
manifold admitting an indecomposable but non-irreducible holonomy
representation, (i.e., with a one-dimensional invariant lightlike
subspace) is a degenerate Kundt (degenerately reducible) spacetime
\cite{kundt}, which
contains the VSI and (non locally homogeneous) CSI subclasses (in which all of
the scalar invariants are zero or constant, respectively) as special cases
\cite{invariants}.} is not completely classified by its set of scalar polynomial curvature
invariants \cite{invariants}. It is perhaps within string
theory that the full richness of Lorentzian geometry is realized,
where the Kundt spacetimes may play a fundamental role \cite{universal}.

Assuming the existence of two levels -- microscopic and
bulk\footnote{The word ``macroscopic'' would be used here in particle
physics. However, in the averaging problem in cosmology that we discuss
in the next section there are so many different
scales of averaging that such terminology becomes problematic. Thus we
use the term bulk fields to refer to spatial averages of field theories
involving purely non-gravitational physics.} -- of understanding classical physical phenomena, Lorentz formulated
a microscopic theory of electromagnetism and showed Maxwell's
theory to be its bulk version \cite{Averaging}. A space averaging is always necessary and
unavoidable in all settings that deal with bulk matter fields.
However, in electrodynamics the
field operator is linear in the fields and it can be easily averaged, and
models of continuous electromagnetic media which relate to the structure of
averaged (bulk) fields can be constructed. So in the same sense that
(linear) QED is averaged to obtain Maxwell equations, QG should be averaged to get GR. On larger scales, however,
the results of coarse graining or averaging the geometry are expected to be
far from trivial, since the Einstein \FE\ are highly non-linear. We shall
return to this in the next Section.

\section{Large scales: Cosmology}

When one considers the whole Universe, then the open questions about spacetime
topology and boundary conditions become inescapable. Furthermore, if the
\FE\ are to be viewed as dynamical evolution equations that
couple matter and geometry then the split of space and time in any metric
decomposition also becomes a question with physical as well as technical
mathematical implications.

In 1917 Einstein opted for a universe with a $\Rop\times S^3$ closed spatial
topology to avoid the problem of unknown
boundary conditions at spatial infinity \cite{esu}.
Famously, Einstein had to introduce a finely tuned cosmological constant,
$\Lambda$, in order to try to keep the universe static,
when his equations were trying to tell him that a key feature of general
relativity is that its solutions are dynamical.

In Einstein's day the idea that we could
live in a universe where time could have a beginning was simply disregarded
on philosophical grounds. In an eternally existing universe new information
can always be entering our light cone from infinitely far away, posing
a basic problem that completely changes its character if the universe
had a beginning. In the latter case the universe is divided into observable and
unobservable portions, and the question of spatial topology
then becomes inextricably linked with initial conditions and quantum gravity
(if one makes the reasonable assumption that quantum gravity is important
in the very early universe).

The other foundational question about the dynamical coupling of matter and
geometry---the fitting problem \cite{fit} and the scales of applicability
of the \FE---did not stand out as a question in 1917. Given
the prevailing view that nebulae were not separate galaxies at vast distances,
it was reasonable to assume that the Universe existed as a material
continuum consisting of stars with the density of the Milky Way in all
directions. The idea that the energy--momentum tensor was described as dust
coarse grained as stars, and statistically homogeneous, was observationally
justified.

One hundred years on, our observations of the late epoch Universe reveal a
significantly more complex picture, however. Stars and black holes form
galaxies of a wide range of sizes, while groups and clusters of galaxies
form the largest gravitationally bound structures. These structures themselves
form knots, filaments and sheets that thread and surround very underdense
voids, creating a vast cosmic web \cite{web}. Some 40\% of the volume of the
present Universe is in voids of just one characteristic diameter \cite{HV1},
$\goesas30\hm$, and density contrast $\delta_\rho=(\rho-\bar\rho)/\bar\rho
<-0.94$ which are close to being empty ($\delta_\rho=-1$). Once the distribution
of all voids is accounted for, then by volume the present universe is
void--dominated \cite{Pan11}.

The {\em fitting problem} \cite{fit} -- namely how does one coarse grain
matter and geometry on a given scale to fit it into an effective geometry on
larger scales -- is perhaps the most important unsolved problem\footnote{The
importance of this problem was first recognized by Einstein and Straus in
their Swiss cheese model in 1945 \cite{ES}. At
that time the existence of many of the structures in the
coarse graining hierarchy (\ref{coarse}) was still unknown, and they replaced
it by the simpler scheme $\metric{stellar}\to\metric{universe}$, wherein
a spatially homogeneous FLRW
model is {\em assumed}, eliminating the possibility of backreaction
on average expansion but allowing potential differences for light propagation as
compared to the FLRW case.}
in mathematical cosmology. The observed complex lumpy
universe demands a hierarchy of steps in coarse graining \cite{dust},
which can be depicted as:
\beq
\left. \begin{array}{r}
\metric{stellar}\to\metric{galaxy}\to\metric{cluster}\to
\metric{wall}\\ \vdots\quad \\ \metric{void} \end{array}
\right\}\to \metric{universe}
\label{coarse}\eeq
if we assume a metric description. At one extreme we have discrete particles,
ranging in size from isolated electrons
and protons that fill the vastness of voids, where bound structures never
formed, to stars and supermassive black holes that are the basic building blocks
of more complex gravitationally bound systems.
In the standard FLRW cosmology it is implicitly assumed that regardless
of the gravitational physics in the coarse graining hierarchy (\ref{coarse}),
at the final step\footnote{The smallest scale on which a notion of statistical
homogeneity arises is $70$--$120\hm$ \cite{sdb12}, based on the two-point
galaxy correlation function. However, variations of the number density of
galaxies of the order 7--8\% are still seen when sampling on the largest
possible survey volumes \cite{h05,sl09}.} the matter distribution
can be approximated by an ``effective averaged out'' stress-energy tensor.
Moreover, the averaged stress-energy is assumed to be spatially homogeneous with
dust equation of state, and to satisfy the \FE\ with a cosmological constant.
Such an assumption is simply not justified from first principles, however.

The averaging of the Einstein \FE\ for local inhomogeneities on
small scales can in general lead to very significant dynamical
effects---{\em backreaction}---on the average evolution of the
Universe \cite{bu00}. Furthermore, averaging (and inhomogeneities in general)
can affect the interpretation of cosmological data \cite{BC,clocks,AAC2}.

Almost all deductions about cosmology are based on null geodesics: light paths
that traverse the greatest accessible distances. However, inhomogeneities bend
null geodesics and can drastically alter
observed distances when they are a sizeable fraction of the curvature radius. In
the real Universe, voids occupy a much larger volume as compared to bound
structures. Hence light preferentially travels much more through underdense
regions and the effects of inhomogeneities on luminosity distances
can be significant.

One further problem of interpreting observations is that it is necessary, in
principle, to model properties of (not only a single photon) but of a `narrow'
beam of photons. Since the optical scalar equations \cite{opt} (which are
non-linear and follow the geometric optics approximation) require integration
along the beam of null geodesic congruences within a lumpy matter distribution,
there may be important resulting averaged effects. Thus it is also of
importance to study the effect of averaging on a beam of  photons in the
geometric optics limit. The small scale lensing effects of a lumpy matter
distribution in an underdense universe principally involve Weyl curvature.
However, since the FLRW geometry has zero Weyl curvature, if it is to provide
an accurate effective description on cosmological scales, then the averaging of
the optical equations must somehow replace the Weyl curvature due to
isolated masses by a larger scale averaged Ricci curvature.

A theoretically conservative approach is to assume that GR is a (classical)
{\em mesoscopic}
theory\footnote{Since GR incorporates the Strong Equivalence Principle, it
already includes a relevant scale -- local inertial frames -- for the
``microscopic physics'' incorporating all the non-gravitational forces of
the matter sector. Our use of the word ``mesoscopic'' here refers to scales
possibly up to galactic scales, which is, of course, considerably larger than
those envisaged with the use of this terminology in condensed matter physics.}
applicable on those small scales on which it has actually
been tested, with a local metric field (the geometry) and matter fields.
In this idealization, while
strong field gravitational physics is required in the interaction of black
holes, for the most part real particles are modelled
as point particles moving along timelike geodesics in the absence of external
forces, and photons move on null geodesics.

After coarse graining\footnote{The terms ``coarse graining'' and ``averaging''
are often used interchangeably, and should be understood as employing bottom-up
versus top-down approaches to the same problem. Given the historical
importance of the FLRW models, the top-down approach has been much more
widely studied.} we obtain a smoothed out macroscopic geometry (with
macroscopic metric) and macroscopic matter fields, valid on larger scales.
In fact, (\ref{coarse}) indicates a succession of macroscopic scales.
A photon follows a null geodesic in the local geometry. But what trajectories
do photons follow in the averaged macro-geometry? The averaged
vector is not necessarily null, geodesic (or affinely parametrized) in the
macro-geometry \cite{AAC2}. This will
affect cosmological observations. Similarly,
the averaged matter does not necessarily move on timelike geodesics of
the averaged metric. After all, in the final steps of the hierarchy
(\ref{coarse}) we are no longer dealing with particles, but ``fluid'' cells.

The coarse grained or averaged \FE\ need not take the same mathematical
form as the original \FE. Indeed, in the
case of Zalaletdinov's Macroscopic Gravity approach \cite{Averaging} the
averaged spacetime is not necessarily even Riemannian. A rigorous mathematical
definition of averaging in a cosmological model is necessary. Averaging
often involves replacing variables by average values after integration
over a domain. This is possible for scalars \cite{bu00}, but is
problematic for non-scalars unless one introduces additional mathematical
structures \cite{Averaging}. Numerous mathematically consistent approaches to averaging are
possible, but any physically well-motivated approach should be both
consistent
with observations and with the principles of GR. Ideally,
it should extend those principles in the most minimal fashion that
seeks to understand open questions such as the nature of gravitational
energy.

One profound consequence of the Strong Equivalence Principle is that
gravitational energy cannot be localized at a point, but instead is
non-local. The dynamical non-linear coupling of matter and geometry
is therefore crucial to the definition of the ``rest energy'' assigned
to regions at any level of the coarse graining hierarchy (\ref{coarse}).
It is also crucial to defining the relationship between the geodesic
of any observer at the microscopic level of stars and the average
effective geodesic of a coarse grained cosmic fluid cell.

In the standard approach to cosmology one implicitly assumes that
irrespective of the dynamical coupling of matter and geometry, each step
of coarse graining (\ref{coarse}) will determine the velocity of one particle
relative to the centre of mass of the coarse-grained particle at the next
macroscopic scale by a pointlike boost, so that the whole succession of steps
amount to a single boost with respect to the one global FLRW frame, explicable
by an effective Newtonian gravitational potential.\setcounter{footnote}{0}

Nothing in GR demands such an outcome from the physics of the unsolved fitting
problem\footnote{When Einstein's equations are applied to fluids, rather than
to fundamental fields, one is assuming a notion of coarse graining of matter
that is well-established in non-gravitational physics. In standard kinetic
theory one coarse grains by {\em filtering}, i.e., neglecting stochastic
fluctuations in phase space variables in favour of a mean field description.
This is well understood for particles in the absence of spacetime curvature,
but becomes an altogether different problem when coarse graining geometry.}.
Furthermore, since the regional coarse graining of quasilocal gravitational
energy is necessarily involved, we are talking about the very problem -- the
cosmological energy budget -- for which unknown sources of dark matter and
dark energy are added in the standard model of cosmology to make gravity
stronger than our na\"{\i}ve expectation on the level of bound structures
on one hand, and weaker than our na\"{\i}ve expectation on the larger scale of
unbound expanding structures on the other.

Setting aside the evidence of the CMB, the FLRW geometry is generally
invoked ``because it works''. The fact that $\Lambda$CDM works
so well, with just two essential parameters characterizing
unknown physics, suggests that there are simplifying principles to be
found in the unsolved gravitational physics of the fitting problem. While
it is not our intention to debate the merits of particular proposals here,
we point out that the timescape scenario \cite{clocks,rad}
attempts to explain how observed cosmic expansion is so close to uniform
despite late epoch inhomogeneity, by applying a simplifying principle---the
Cosmological Equivalence Principle \cite{cep}---that extends the
Strong Equivalence Principle to cosmological averages. It results in
a phenomenological description which is competitive with the $\Lambda$CDM
model in independent tests that are currently possible \cite{tstest},
and it makes concrete predictions \cite{obs} for future tests that can
distinguish it from $\Lambda$CDM, including the
redshift time drift test \cite{red} and the Clarkson--Bassett--Lu (CBL) test
\cite{cbl,smn}.

The Euclid satellite will enable an extremely precise test of the
validity of the FLRW geometry by the CBL test according to recent estimates
\cite{smn}, and also by the distance--sum--rule test \cite{rbf}. Given
the prospect of such tests in the next decade, and the
fact that the fitting problem involves some of the deepest unsolved
foundational questions in GR, we suggest that real progress can be made
in the coming decades if researchers tackle the fundamental principles of the
fitting problem while thinking hard about new data.

\section{Large scales meet small scales: the very early Universe}

The foundational questions faced by quantum theory and by cosmology become
inextricably entangled in the very early Universe. The isotropy of the CMB
on all scales larger than a degree indicates a prior state of thermal
equilibrium between regions which cannot have been in causal contact given
the expansion history of FLRW universes containing just non-relativistic
matter and radiation. This is the {\em horizon problem}. The problem
it poses for causal structure is one of the key reasons that the inflationary
paradigm was introduced almost 40 years ago. Generically, with additional
matter degrees of freedom, such as scalar fields, a very early period
of de Sitter--like exponential expansion is produced, leading to vast
changes in the structure of the past light cone. Inflation thereby
not only resolves the horizon (and flatness) problems, but
via quantum effects the energy of
the new fundamental fields is converted to create all the particles of
the Universe when inflation ends.

The phenomenology of inflation has been extraordinarily successful in
accounting for a nearly scale invariant spectrum of density perturbations
which in turn give rise to temperature fluctuations on the last scattering
surface, with an anisotropy
spectrum that well matches what is observed. Yet inflation
remains a phenomenology, in search of a fundamental theory. Whether a
given model inflates or not, and by how much, often depends on initial
conditions, effectively pushing the foundational questions back to
those of quantum gravity.
Moreover, many models of inflation are now beginning to be ruled out by
observations from the Planck satellite \cite{M16}: in particular, those that
give rise to the production of copious primordial gravitational
waves. The combined Planck/BICEP2/Keck observations \cite{BK}
now yield an upper bound on the ratio of tensor to scalar power of
$r<0.07$ at 95\% confidence, (with a pivot scale $0.05\,\hbox{Mpc}^{-1}$).
Typical models which survive are single field inflationary models with a long
plateau, similar to the Starobinsky model \cite{S80}, based on a $R+R^2$
extension of minimal Einstein gravity. Such models may naturally include ``no
scale'' supergravity compactifications \cite{ns}, or the Higgs boson itself as
the inflaton provided it has a strong non-minimal coupling to gravity
\cite{bs}, which would pose challenges to standard field-theoretic approaches
for the unification of gravity with gauge interactions \cite{ns}.

Since the Higgs mechanism is associated with phase transitions in which
standard model gauge bosons first acquire masses, from the first principles of
GR it is natural to ask whether the simplistic picture of spacetime
structure that underlies the conceptual framework of inflation  is, in fact,
the correct picture. In particular, does a single ($1+3$)--dimensional
classical
manifold (or a $1+3+6$ or $1+3+7$ higher-dimensional manifold) somehow
nucleate at the Planck scale, giving us quantum field theoretic physics based
on a fixed spacetime? (This is the picture that underlies most current very
early Universe frameworks.)
Or should the conventional notion of spacetime structure itself emerge
at energy scales lower than the Planck scale?

After all, the Planck scale is
derived by simple dimensional analysis extrapolated from the known
fundamental constants.
It therefore represents a limit in which we know our understanding of
spacetime structure breaks down. However, between the electroweak scale
and the Planck scale we simply extrapolate the idea that spacetime
structure must be fixed, because that is the physics we are familiar with.
All of this could change, in ways which change the effective couplings between
gravity and the Higgs sector from the field theoretic view, if the problem has
a more fundamental origin involving the dynamical coupling of matter and
geometry at the quantum level.

It is remarkable that many independent approaches to quantum
gravity seem to find an effective ``spontaneous reduction to two dimensions''
at high energies somewhat below the Planck scale \cite{C12}.
Such approaches include causal dynamical triangulations \cite{cdt}, renormalization group methods
\cite{rg} which invoke asymptotic safety, and loop quantum gravity \cite{lqg},
among others.
Since one is dealing with an effective reduction of the spectral dimension,
or the dimension of the space probed by typical geodesics, a ``reduction
to two dimensions'' should be pictured in terms of an inherently quantum
$(2+2)$--dimensional phase space rather than classical particles
in a 2-dimensional configuration space.

As Carlip points out \cite{C12}, the short-distance Wheeler-DeWitt equation may
be dominated by ``asymptotically silent spacetimes''
in which light cones shrink to lines and nearby points become causally
disconnected. This leads to Belinsky--Khalatnikov--Lifschitz (BKL) asymptotics \cite{bkl}, in which the metric
is locally Kasner with axes of anisotropy that vary chaotically. Each point
effectively has a ``preferred'' spatial direction and geodesics effectively
see only $1+1$ dimensions \cite{C12}. If one has asymptotic silence everywhere,
then the small scale metric will have two length scales
\beq
ds^2 = \ell^2_\parallel g_{\alpha\beta}dx^\alpha dx^\beta +\ell^2_\perp
h_{ij}dx^i dx^j\label{d2}
\eeq
and the Einstein-Hilbert action becomes very nearly the action of a
2-dimensional conformal
field theory for the transverse metric $h_{ij}$ \cite{C12}. Such $2+2$
decompositions (\ref{d2}) have previously been studied in very high--energy
gravitational scattering processes from a field theoretic viewpoint \cite{tvv},
and also give a useful framework for investigating quasilocal energy exchange
in classical GR \cite{u16}.

The spontaneous dimensional reduction to two dimensions \cite{C12} of course
encompasses ideas which are still speculative, and not fully understood. But it
illustrates a convergence between the geometric and field theoretic approaches
to quantum gravity, and the phenomenology of the early Universe. In particular,
we have an effective scale invariance at very high energies, and processes
such as those based on the metric (\ref{d2}) which formally at least resemble
those of string theory \cite{tvv}.

To make progress in the mathematical modelling of such ideas we need to change
the conceptual picture of spacetime as a fixed manifold in which particles
live as classical or quantum fields. The essence of GR is that spacetime is
a relational structure between material particles.
While our conventional picture works
at low energies, at very high energies when the average relationship between
particles is becoming lightlike, then the relational structure -- i.e., spacetime
structure -- also appears to change in all the quantum gravity approaches that
see an effective reduction to two dimensions.

Time is a property only measured by massive timelike particles undergoing the
physical processes of particle physics that give rise to reaction rates,
scattering amplitudes, and binding to form condensates. Massless particles do
not carry clocks. The fact that successful inflationary phenomenologies are
leading to models which are ``scale free'' in some sense, or which require
unconventional non-minimal coupling between gravity and the Higgs sector,
suggests that the high energy transition from a universe containing only
massless particles to one in which particles acquire masses could also
involve a phase transition in spacetime structure.

Such a scenario may appear radical when compared to current theories
which generally invoke phase transitions in the very early Universe as occurring
within field theories on a global FLRW spacetime. However, it is closer
to the basic innovation of GR that matter and geometry are dynamically
coupled. Moreover,
it could lead to a deeper theoretical basis for explaining the phenomenology
of inflation. The horizon problem is solved via a transition from a simpler
asymptotically silent effectively 2--dimensional initial quantum state. The
statistical spatial homogeneity and isotropy of the present epoch Universe
could then be a consequence of a degree of scale invariance that survives
from the very early Universe in the average geometry of the present Universe.
A physical principle such as the Cosmological Equivalence Principle \cite{cep}
may then be a constraining principle for relevant mathematical structures
that relate the effective geometry of the early and late epoch Universes,
without invoking a global FLRW geometry on the largest scales.

This paper is intended to initiate discussion about these questions,
rather than advocating any particular approach. The questions are timely
since current cosmological observations
not only throw up fundamental mysteries such as dark energy and dark matter, but are now
also strongly constraining models
of inflation \cite{M16}. What survives are phenomenologies \cite{ns,bs}
that have
 a strong resonance with findings from many different approaches to quantum
gravity \cite{C12}, and with the idea that gravity emerges as a breaking
of conformal invariance \cite{cbreak}. There have already been studies of
dynamical dimensional reduction in the early Universe, in terms of the effect
of modified dispersion relations on cosmological perturbations \cite{gm}.

It is, however, high time to rethink the mechanisms by
which the spacetime of GR emerges from quantum
gravity in the early Universe. String theory is already built on the conformal
invariance of the 2-dimensional string worldsheet. However, it treats spacetime
as a separate entity: a unique higher-dimensional
continuum chosen somehow from a multiverse,
a statistical collection of vacua -- geometries of fixed
finite dimension on which matter fields live, classified according to their
(super)symmetries. Loop quantum gravity and other geometric approaches often
focus purely on quantizing geometry, without reference to matter.

Here we offer the view that whatever ends up being retained in a theory
of quantum gravity from either of these approaches, it is likely to more deeply
embody the notion of GR as a theory of dynamical spacetime structure that
{\em couples} matter and geometry at the quantum as well as the
classical level. A logical extension of the principles of GR to a global
theory of spacetime structure may well mean that while the
Einstein \FE\ hold at small scales at late epochs, there is no
need for these equations to hold on all scales at all epochs. Rather
than having a single notion of geometry on all scales while modifying
the Einstein-Hilbert action, we might retain the Einstein-Hilbert action
on mesoscopic scales but change the notion of the geometrical structure of
spacetime at early epochs, and on large scales. Rather than dealing
with Modified Gravity we are then dealing with {\em Modified Geometry}.

GR is not a finished theory that was completely solved
100 years ago. The next 100 years of the development of relativistic cosmology
are likely to be as exciting as the first 100 years. The problems we
face also demand that we think equally rigorously about the foundations
of GR as Einstein did 100 years before us.\bigskip

\noindent {\bf Acknowledgements}\quad We would like to thank Boud Roukema
for helpful comments. AAC acknowledges the financial support of NSERC.
\appendix
\section{A brief overview of GR}

GR is often described as both
beautiful and elegant. To embody the universal nature of the gravitational
interaction, Einstein re-envisioned gravity as a property of the relational
structure between all matter particles (including massless ones). That
is, gravity is a property of spacetime structure, rather than a force in a
pre-existing spacetime. This distinguishes GR from the theories of
the other fundamental interactions.

In GR it is assumed that all matter moves in an effective
pseudo-Riemannian metric spacetime with a universal coupling, governed by the
Einstein Equivalence Principle, consisting of two parts:
(i) {\em The Weak Equivalence Principle:} Given the same initial positions and
velocities, subject only to gravity particles will follow the same trajectories,
or {\em geodesics}. In other words, particles all fall with the same
acceleration regardless of composition and consequently gravity is universal.
(ii) {\em The Strong Equivalence Principle:} The laws of physics take the
same form in a freely-falling reference frame as in SR.
Effectively, gravity can always be eliminated at a point.
Instead gravity is related to tidal forces -- the deviation
of neighbouring geodesics from the SR expectation.
Measurement of geodesic
deviation, on a scale determined by a suitably large experiment,
allows a direct determination of the Riemann curvature tensor \cite{compass}.

On the road to the Strong Equivalence Principle, in 1907 Einstein formulated
the more limited version that all motions in an external static homogeneous
gravitational field are identical to those in the absence of a gravitational
field if referred to a uniformly accelerated coordinate system. A common
misconception that follows from this is that acceleration requires GR. It
does not; accelerated frames can be treated by general coordinate systems
within SR. As far as GR is concerned, it is the tidal forces that constitute
the true gravitational field.

The Strong Equivalence Principle is satisfied in GR by assuming that all
matter fields are minimally coupled to a single metric tensor,
$g_{\mu\nu}$, with a torsion-free affine connection. Infinitesimal proper
length and time intervals -- those measured by
physical rods and clocks -- between two points (events) in
4-dimensional spacetime are given by the metric line element. On larger
scales proper lengths and times are given by the integral curves
of the geodesic equations, which follow from the line element by a
variational principle. E.g., variation of the proper time, $\tau=\frac1{c}\int
d\lambda\,\sqrt{-g_{\mu\nu}\dot x^\mu\dot x^\nu}$, where $\dot x^\mu\equiv
\frac{d x^\mu}{d\lambda}$ on curves with parameter $\lambda$, leads to the
timelike geodesic equation $U^\nu\nabla_\nu U^\mu=0$, where $\mathbf\nabla$ is
the covariant derivative and
$U^\mu\equiv\frac{d x^\mu}{d\tau}$ is the 4-velocity. Geometrically this
is the equation of parallel transport for the 4-velocity.

The central idea in Einstein's GR is that gravity is described entirely in
terms of the geometry determined from a metric, and that
it is free of any ``prior geometry'' which is fixed immutably and
independently of the distribution of the gravitating sources. Instead, the
metric and the matter sources of energy--momentum in the Universe
are dynamically coupled via
the Einstein field equations, $$G^{\mu\nu}=\frac{8\pi G}{c^4} T^{\mu\nu},$$
where
$G^{\mu\nu}$ is the Einstein tensor which is related to the spacetime curvature
(and is determined by the metric,
$g_{\mu\nu}$, and its first and second partial derivatives), $G$ is Newton's constant, and
$T^{\mu\nu}$ is the energy-momentum tensor.
The \FE\ can also be obtained in general from
the Einstein-Hilbert action:
$$S_{\rm grav}=\frac{c^{3}}{16\pi G}\int d^{4}x \sqrt{-g}\,{\cal R},$$ where
${\cal R}$ is the Ricci scalar curvature, plus a minimally coupled
action, $S_{\rm matter}$, for the matter fields on small scales.
However, on mesoscopic scales the energy-momentum tensor is already assumed
to be averaged as an effective fluid rather than being directly derived from
an action.

There have been many successes of GR, from the predictions of the motion of
planets, satellites (and GPS), the deflection of light and the Shapiro
time delay in the solar system, to numerous strong field processes
in astrophysics and cosmology, including in particular
the prediction of black holes. Over the past century GR has been accurately
tested at scales from $10^{-4}\,$m in laboratories, up to $10^{14}\,$m in
the solar system and in strongly gravitating binary pulsar systems
\cite{Will93}. Perhaps, most excitingly, the first
direct observation of gravitational waves was finally
achieved by LIGO on the 100th anniversary of GR \cite{GW2015}.

GR remains untested at both ends of the spectrum of distance scales,
however. The largest scale on which GR is directly tested is still 12 orders
of magnitude smaller than the present size of the observable
Universe, $\goesas10^{26}\,$m. Furthermore, on small scales gravitational
effects are difficult to test directly due to the weakness of gravity
relative to other fundamental interactions.
Einstein never specified the scale of applicability of
his equations, but given the universal nature of gravity we must assume
that GR applies on all scales unless observational evidence for some
modified theory is found. Quantum mechanics played no role in the original
formulation of GR. However, since quantum mechanics also applies to every form of matter,
we know from simple dimensional considerations that our classical understanding
of spacetime structure must break down at the Planck scale. The development of
a possible quantum theory of GR, and a discussion of the fundamental issues
involved, has certainly garnered much attention. One point that is not often
emphasized is that the Planck scale simply represents the limit of the
``known unknowns''. Quantization of the spacetime manifold itself may
conceivably involve changes to the relations that define spacetime structure 
at scales somewhat larger than the Planck scale.

We may also question whether GR is valid on cosmological scales. Answering
this question observationally is complicated since cosmological measurements
of the evolution of large-scale structure depend
strongly on the underlying cosmological model. Most often this is taken to be
the simple FLRW model based on the assumptions of spatial homogeneity
and isotropy, with the matter content of the \LCDM\ concordance model.
However, this simple model leads to the very problematic issues of
non-baryonic dark matter and dark energy
(such as a cosmological constant, $\Lambda$) which have not been
directly detected, and whose origin remains a mystery.
There is much less discussion of the fundamental aspects of cosmology
\cite{DebonoSmoot,fit,dust} than the fundamental aspects of field theory.
In particular, if the Einstein \FE\ hold on large
scales, they require an effective stress--energy tensor in which
small scale inhomogeneities are averaged out. Any procedure of
coarse-graining involves additional mathematical and physical assumptions,
which in general give rise to {\em backreaction}: namely, average cosmic
evolution that differs from FLRW evolution \cite{BR,buchert11}.

The two regimes in which GR is not yet tested,
the small and large scales, may be linked together
in the very early Universe. For example, in a widely accepted framework
quantum fluctuations just after the Plank regime are stretched by
inflation to produce fluctuations on all spatial scales before the Universe
entered a radiation-dominated regime. While baryons, electrons and photons
remain tightly coupled, the density fluctuations are somewhat amplified on the
small scales determined by the distance that sound waves can propagate.
When the Universe cools enough that the first atoms form, at the last
scattering epoch, the density perturbations are frozen in. The density
perturbations grow via gravitational instability, giving rise to
all the complex hierarchy of structures which are strongly inhomogeneous
on scales smaller than the present comoving scale of the sound horizon,
around $100\hm$, where $h$ is a dimensionless parameter related to
the Hubble constant by $H_0=100\,h\kmsMpc$.

\section*{References}

\end{document}